\documentclass[8.5pt,twoside,twocolumn]{revtex4-1}
\usepackage{graphicx}
\usepackage{xcolor}
\usepackage{hyperref}

\usepackage{amsmath}
\usepackage[binary]{SIunits}
\DeclareMathOperator{\erf}{erf}

\usepackage{algpseudocode}
\usepackage{xcolor}

\begin{document}
\title{A novel particle tracking method with individual particle size measurement and its application to ordering in glassy hard sphere colloids}

\author{Mathieu Leocmach} 
\email{mathieu.leocmach@polytechnique.org}
\altaffiliation[Present address: ]{Laboratoire de Physique, CNRS UMR 5672, Ecole Normale Supérieure de Lyon, 46 allée d'Italie 69364 Lyon cedex 07, France.}
\author{Hajime Tanaka}
\email{tanaka@iis.u-tokyo.ac.jp}
\affiliation{ {Institute of Industrial Science, University of Tokyo, 4-6-1 Komaba, Meguro-ku, Tokyo 153-8505, Japan} }

\begin{abstract}
Particle tracking is a key to single-particle-level confocal microscopy observation of colloidal suspensions, emulsions, and granular matter. The conventional tracking method has not been able to provide accurate information on the size of individual particle. Here we propose a novel method to localise spherical particles of arbitrary relative sizes from either 2D or 3D (confocal) images either in dilute or crowded environment. Moreover this method allows us to estimate the size of each particle reliably. We use this method to analyse local bond orientational ordering in a supercooled polydisperse colloidal suspension as well as the heterogeneous crystallisation induced by a substrate. For the former, we reveal non-trivial couplings of crystal-like bond orientational order and local icosahedral order with the spatial distribution of particle sizes: Crystal-like order tends to form in regions where very small particles are depleted and the slightly smaller size of the central particle stabilizes icosahedral order. For the latter, on the other hand, we found that very small particles are expelled from crystals and accumulated on the growth front of crystals. We emphasize that such information has not been accessible by conventional tracking methods.
\end{abstract}
\maketitle

\section{Introduction}

Experimental physicists often need to recognize objects, count them, follow them and characterise them. \citet{perrin} had to count colloids by hand to establish the sedimentation-diffusion equilibrium. Nowadays computer vision algorithms are used routinely in the lab to track hundreds of thousands of objects as diverse as stars in a galaxy~\cite{Bertin1996}, tracers in a microfluidic device~\cite{Wereley2010}, pattern formation in polymer systems~\cite{tanaka1986application,tanaka1989digital}, dust in a plasma, bacteria~\cite{Zhang2010,Gibiansky2010} or viruses in a living cell~\cite{Brandenburg2007}. In all these cases, tracking is possible if the particles are either point-like and far apart, or several pixel wide and almost monodisperse in size (at least the smallest dimension for anisotropic objects~\cite{Zhang2010,Gibiansky2010}). To our knowledge, algorithm that allow tracking of polydisperse particles in crowded environments have not reached the soft matter community.

Overall particle size distribution in colloidal suspensions and emulsions influences crystallisation~\cite{Pusey1987,Henderson1996,Fasolo2003,Schope2007,pusey2009hard}, glass forming ability~\cite{Pusey1987,Henderson1996,Senkov2001,Schope2007,pusey2009hard}, sedimentation~\cite{Binks1998,Leocmach2010} and emulsion stability~\cite{Biben1993,Binks1998} among other physical phenomena. It can be characterized by various methods that rely on measurements done in well-controlled environments~\cite{Lange1995,Provder1997,Finder2004}. However the local size distribution is not accessible experimentally \emph{in situ} and has thus not been studied so far.

Particle-level microscopy experiments usually access the coordinates of the particles via the algorithm proposed by \citet{Crocker1996}. The original noisy image is blurred by convolution with a Gaussian kernel of width $\sigma$ to yield a soft peak per particle. Local intensity maxima within this blurred image give the coordinates of the particles with pixel resolution. Sub-pixel resolution ($0.1\sim0.3$~pixels error) can be achieved by taking the centre of mass of a neighbourhood around the local maxima. The extension of this algorithm to localize particles in three-dimensional (3D) confocal microscopy images has been done in two ways: either tracking particles in each confocal plane and reconstructing the results (2D-flavour)~\citep{vanblaaderen1995rss, Lu2007}, or full image analysis on three dimensional pictures (3D-flavour)~\citep{dinsmore2001tdc}.

The choice of the width $\sigma$ of the blurring kernel is critical: if it is too small, then the intensity profile is flat near the centre  of a particle, leading to multiple and ill-localized maxima per particle; if it is too large, then the peaks of nearby particles overlap, leading to shifts in the detected positions~\citep{Baumgartl2005,Jenkins2008}, or even fusion of the particles (only one particle detected instead of two). If the colloids are fairly monodisperse one can argue (at least in the 3D-flavour) that there exists a range of possible width where the choice of $\sigma$ has almost no effect on the number of particles detected. Choosing $\sigma$ within this range gives confidence in the localisation results.

\begin{figure*}
\centering
\includegraphics{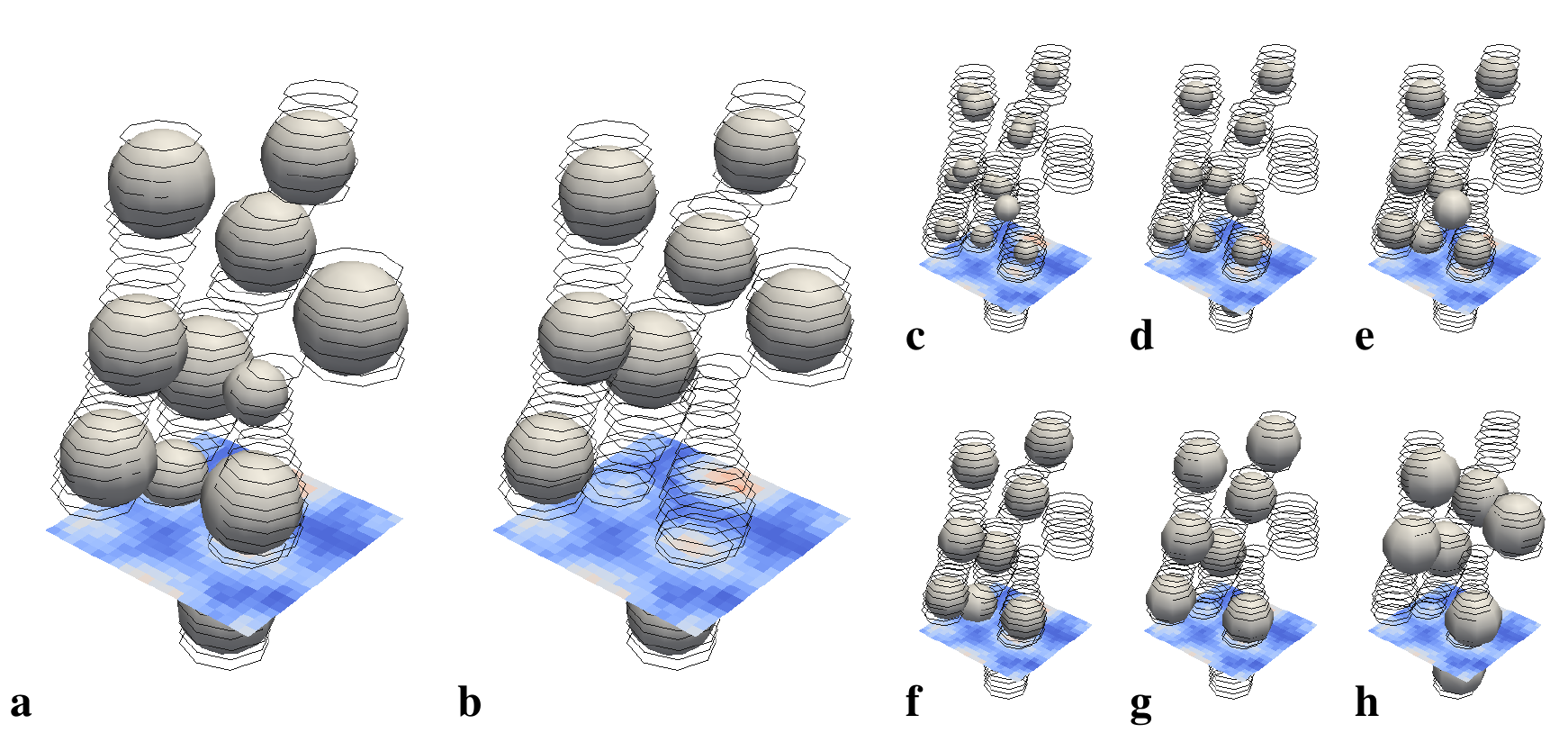}
\caption{Visualisation of the results of various tracking methods for the same portion of image. (a) Multiscale 3D tracking. (b) Reconstruction from multiscale 2D tracking. (c-h)  Crocker and Grier method in 3D with blurring radius increasing from \unit{2}{px} to \unit{4.5}{px} by steps of \unit{0.5}{px}. The circles on each picture are the result of 2D multiscale tracking of each XY slice of the 3D pictures. Sphere are displayed with radii determined by the tracking methods in (a) and (b), and equal to the blurring radius for (c)-(h).}
	\label{fig:localise}
\end{figure*}

However, we found that no such ``good blur width'' exists in a sample of moderate ($6-7\%$) polydispersity (see Fig.~\ref{fig:localise}c-h). The detection of smaller particles with small blurring is incompatible with the detection of the larger particles, and conversely. This unacceptable failure of the \citet{Crocker1996} algorithm, as well as the want of the particles' radii data, triggered our design of a novel localisation algorithm that would be robust even for a system of any finite polydispersity, which is unavoidable and sometime desired in real experiments. Recently \citet{Kurita2011,Kurita2011b} have designed a sizing method using particle coordinates from confocal experiments. However their method do not work at the image processing level and relies on coordinates extracted via the \citet{Crocker1996} algorithm. If these coordinates are wrong or if some particles are missed (as shown in Fig. \ref{fig:localise}), the output of their method could not be exact.

The key notion to detect objects of unknown and possibly diverse sizes in an image is the \emph{scale space}~\cite{Lindeberg1993}. A popular implementation for isotropic objects (or ``blobs'') is the Scale Invariant Feature Transform (\textsc{sift}) of \citet{Lowe2004}. It is often used to match between different images from complex objects consisting of many rigidly linked blobs (\emph{e.g.}, to create a large scale image from overlapping pictures)~\citep{Lowe2004, Urschler2006, Cheung2009}. To our knowledge, this method has never been used for the quantitative localisation and sizing of independent single-blob objects like spherical colloids, droplets in an emulsion, or crystal nuclei. 

Here we apply this new particle tracking method to study how the sizes of particles affect local structural ordering 
in a supercooled colloidal suspension and the process of heterogeneous nucleation from a substrate. 
We reveal non-trivial local couplings between such orderings and the spatial distribution of particle sizes, which may provide crucial information for our understanding 
on how the polydispersity influences liquid dynamics and ordering phenomena.    

The organization of our paper is as follows. 
In Section~\ref{sec:method}, we will describe our localisation method and its results on synthetic but more and more realistic data. In Section~\ref{sec:confocal} of the paper, we will focus on the specific case of 3D confocal data. In Section~\ref{sec:yon6} we will apply our method to a crystallizing glass of polydisperse hard spheres observed by confocal microscopy and discuss the influence of size distribution on the ordering process. We conclude in Section \ref{sec:conclusion}. 

\section{Localisation and sizing method}
\label{sec:method}

In this section, we will start by recalling the principle of \textsc{sift}, then we will explain our method in the ideal case of an isolated binary ball, to add successively finite dilution and difference in brightness between the particles.

\subsection{Scale invariant feature transform} \label{sec:blur}
The \textsc{sift} consists in convolving the original image $I$ by Gaussian kernels $g_{\sigma_s}$ of logarithmically increasing widths $\sigma_s$ to obtain a series of blurred images $G_s$
\begin{align}
\forall s>0,\quad G_s = I \star g_{\sigma_s}, \label{eq:gaussian_blur}\\
\intertext{where $\star$ is the convolution operator and}
\forall s>0,\quad \sigma_{s} = 2^{s/n} \sigma_0 ,
\label{eq:sigma_s}
\end{align}
with $n$ a fixed integer. Following Ref. \cite{Lowe2004} we use $\sigma_0=1.6$ and $n=3$. Bright objects in the original image appear as bright blobs in the blurred images, and the blobs fuse together as the kernel width increases (see Fig.~\ref{fig:localise}c-h). This can be seen as a series of low-pass filters in the frequency domain. If we take the difference between consecutive blurred images we obtain a comb of band pass filtered versions of $I$: 
\begin{equation}
\forall s>1,\quad DoG_s = G_s - G_{s-1} \label{eq:DoG_s}
\end{equation}
The difference of Gaussians ($DoG$) response function defined in this way depends on the position in space $\vec{r}$ and on the scale $s$. Bright objects in the original image are detected as local minima in $DoG$. With this procedure any feature with a radius as small as \unit{2}{px} and as large as its distance to the edges of the image can be detected. Furthermore, the intensity of the response is optimal at a $\sigma$ that can be related to the size of the object (see below). Thus a local minima in both space and scale in the response function $DoG$ indicates both localisation and \emph{size} of an object, without any assumption on the target size.

\begin{figure}
	\centering
	\includegraphics{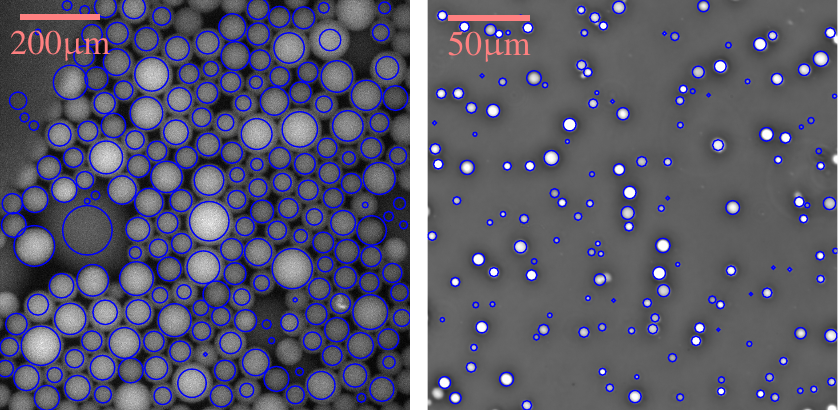}
	\caption{Application of \textsc{sift} in 2D. (a) Fluorescent droplets in a microfluidic device. The wall of the device is visible in the top-left corner. 
	(b) Nucleation under phase contrast microscope. The blue circles are the result of our algorithm. For any tracking algorithm some detection failure near the edge of the image are unavoidable and indeed visible here since we show the whole pictures used for tracking.}
	\label{fig:applications}
\end{figure}

Because of the inherent polydispersity of many soft matter systems, the possible applications of \textsc{sift}-based localisation are countless. For example, droplets of an emulsion can be followed through a microfluidic device. Fig.~\ref{fig:applications}a shows the result of our version of the \textsc{sift} on a very polydispered and dense emulsion observed by wide-field fluorescence microscopy~\cite{Montagne2011}. The sizes extracted are obviously correct except when the system cannot be considered as 2D. Another possible application is nucleation rate measurement. When a phase transition proceeds via nucleation growth, the sizes of the nuclei are very diverse, making their automatic counting difficult by other methods. As shown in Fig.~\ref{fig:applications}b the \textsc{sift} allows to count and to measure the size of nuclei during a liquid-liquid transition in a water-glycerol mixture~\cite{Murata2012}.

However, to the extent of our knowledge, the \textsc{sift} has not been used in a physics context to obtain reliable measurements of particles numbers and sizes from well calibrated images. We will address this reliability in the following.

We will mainly cope with the three-dimensional extension of the \textsc{sift}. As in the case of the \citet{Crocker1996} algorithm, the \textsc{sift} can be extended in three dimensions (\emph{e.g.}, confocal microscopy images) in two ways: either by extracting 2D blobs independently on each slice and then reconstructing 3D objects (Fig.~\ref{fig:localise}b) or by working directly in three dimensions~\cite{Urschler2006, Cheung2009} (Fig.~\ref{fig:localise}a). We found that the former method is prone to errors, missing about a tenth of the particles in our best implementation (Fig.~\ref{fig:localise}b). The 3D results presented in this paper are obtained solely by the later method.

We stress that a volumetric implementation of \textsc{sift} implies a large amount of data (typically more than \unit{1}{\giga b} for a $(\unit{256}{px})^3$ picture) and thus requires careful memory management. Our best implementation (C++) on a 4 cores i7 computer takes less that \unit{10}{\second} to extract the positions and scales of $\sim 10^4$ particles in such a volumetric picture. A much slower implementation (Python+Scipy, single core) takes \unit{1}{\second} to deal with $(\unit{1600}{px})^2$ 2D images like Fig.~\ref{fig:applications}b. We would expect real-time processing for 2D images on GPU-enabled implementations.

\subsection{Sub-pixel and sub-scale resolution}

\begin{figure}
\centering
\includegraphics{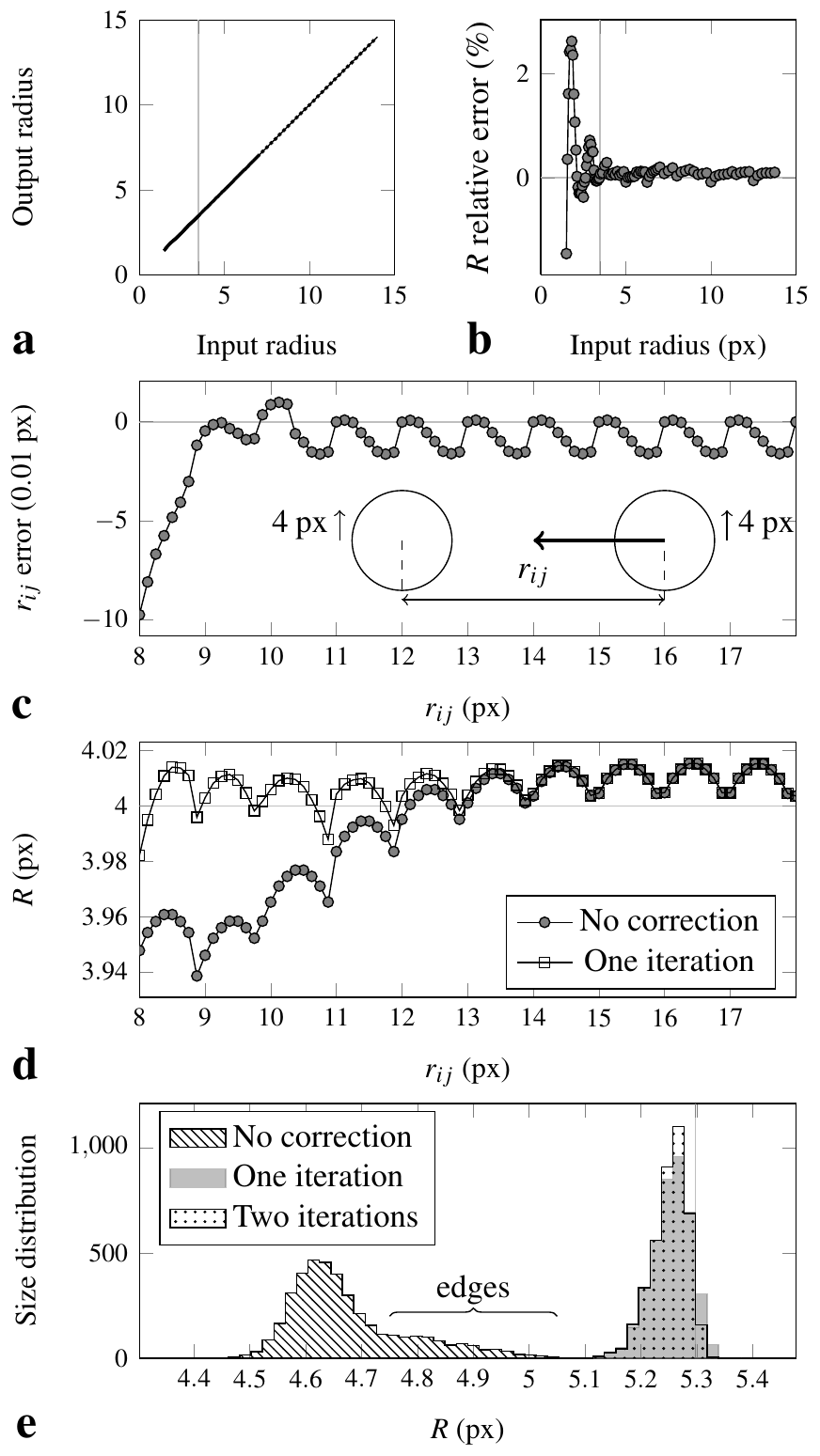}
	\caption{Results from perfect images. (a-b) Sizing of an isolated sphere. Left of the vertical line our algorithm uses a doubled image. (c) Localisation error and (d) sizing error function of the distance between two particles. Oscillations are due to off-lattice centre position. (e) Size distribution extracted from digitized configuration of 4000 monodisperse hard spheres at $0.50$ volume fraction (the vertical line indicates the input radius). The tail to the right is due to particles on the edges of the image who have fewer neighbours and thus are more `dilute'. (d) and (e) also show the effect of finite dilution correction up to convergence.}
	\label{fig:perfect}
\end{figure}

Here we assume perfect noiseless, distortion-free, images. The objects to localise are thus (pixellised) balls of uniform intensity. To mimic the low resolution of experimental images in our test images (see Fig.~\ref{fig:perfect}), we draw uniformly white balls on a 4 to 16 times larger image and then we reduce accordingly the resolution using area resampling.

The $DoG$ constructed above is defined on a $d+1$ dimensional grid, with $d$ the spatial dimensionality. One can interpolate it as a continuous function of the position $\vec{r}$ and the scale $s$ and thus localise the minima with a precision below the grid size, which corresponds to sub-pixel resolution on the position and sub-scale sizing~\citep{Lowe2004}. We found that second order estimate of the spatial derivatives and first order estimates of the scale derivatives gave the best precision.

The object-by-object optimal scale determination allows us to perform the spatial sub-pixel resolution step for each object on an image that is blurred just enough to have neither a flat intensity profile nor an influence of the overlap with a nearby object's blob (an effect that plagues the \citet{Crocker1996} algorithm~\cite{Baumgartl2005,Jenkins2008}). We found that if for a given object the $DoG$ is minimum at $\sigma_s$, the best image to use is $G_{s-1}$. This leads to a spatial resolution below $0.1$~pixels in the worst case (when particles are at hard-core contact), which is the same as the average precision claimed by \citet{Crocker1996}. Moreover, when particle's surfaces are further than $1$~pixel, the error on the positions is less than $0.02$~pixels ($0.3\%$ of the diameter) (see Fig.~\ref{fig:perfect}c).

\subsection{Sizing at infinite dilution}
\label{sec:dilute}
The analytical response of a binary ball to a Gaussian blur, at the centre of the ball is a function of dimensionless ratio $x=R/\sqrt{2}\sigma$: $G(x)$ (see Appendix for an exact expression). The $DoG$ response function is the difference between the two such functions. However, the choice of the width of the two functions is not arbitrary: we make the difference between two consecutive blurred images, the image blurred by $\sigma_{s+1}$ and the image blurred by $\sigma_s$. Therefore, in each of our discrete $DoG$ images the value at the centre of the particle can be expressed as
\begin{align}
DoG(R,\sigma_s, \alpha) = DoG(x_s, \alpha) &= G(x_s/\alpha) - G(x_s),\\
\intertext{with $\alpha=2^{1/n}$. Given sub-scale refinement, this can be written as a continuous function of $\sigma$:}
DoG(R,\sigma, \alpha) = DoG(x, \alpha) &= G(x/\alpha) - G(x).
\end{align}

Here it is clear that minimizing $DoG(x, \alpha)$ with respect to $x$ yields a value $x^*$ that depends only on $\alpha$. Exact calculation yields (see Appendix):
\begin{equation}
	x^* = R/\sqrt{2}\sigma^* = \sqrt{\frac{d\ln \alpha}{1-\alpha^{-2}}}, 
	\label{eq:scale_dil}
\end{equation}
where $d$ is the spatial dimensionality. 
Practically one obtains $\sigma^*$, the value of $\sigma$ that minimises $DoG(R,\sigma, \alpha)$, by a polynomial fit for discrete data $\sigma_j$. Eq.~(\ref{eq:scale_dil}) allows to translate the $n$-dependent $\sigma^*$ to the parameter-free real radius of the particle, $R$. The error on $R$ does depend on the number of subdivisions. We found that with $n=3$ the radius of an isolated pixelated ball can be indeed measured within $0.3\%$ relative error with this method (see Fig.~\ref{fig:perfect}b).

The scale $s=0.5$ corresponds (via Eq.~(\ref{eq:sigma_s}) and Eq.~(\ref{eq:scale_dil})) to the smaller detectable radius $R_{min}\approx \unit{3.5}{px}$ ($\sigma_0=1.6$, $n=3$). In order to detect small objects, \citet{Lowe2004} recommends to double the size of the input image using linear interpolation prior to building the first level of the pyramid. This method fares relatively well in noiseless images (see Fig.~\ref{fig:perfect}b) despite larger errors, but we found that it has to be used with care in confocal microscopy images due to noise and deconvolution artefacts. In addition doubling an image size implies a 8-fold increase in memory consumption (reaching \unit{60}{\giga\byte} in the case of a $(\unit{512}{px})^3$ original image). Our implementation allows this on $\unit{64}{\bit}$ computers by relying on memory mapped files, nevertheless it is mostly impractical. All following tests results are obtained without relying on doubled images.

\subsection{Edge and overlap removal}

The difference-of-Gaussian function has a strong response not only at the centre of bright objects but also along their edges. To eliminate these spurious detections, \citet{Lowe2004} suggests to construct the local Hessian matrix around each minimum of the $DoG$ and then compare its eigenvalues to identify elongated objects.

We found that this was often not enough, especially in crowded environments where an isolated void induces local minima of the $DoG$ response with rather isotropic signatures on the edge of nearby particles. An other case not covered by the Hessian technique is the physically hierarchical structure of many soft matter systems, \emph{e.g.}, particles forming clusters. In this situation our algorithm detects a blob for each particle and a much larger blob for the cluster. In both cases the $DoG$ response of the spurious feature is smaller (less negative) than the one of the valid particles covering the same portion of space. 

Our method to remove the spurious feature is as follow: we looked for pairs of particles closer to each other than the larger of their radii (assuming infinite dilution). This means the centre of one of the particles is situated inside the other. In the physical systems we study, this cannot be correct due to excluded volume effects, thus we remove the feature of lesser $DoG$ response. We name this method `half overlap removal'. One may be tempted to implement a full overlap removal (no pair of particle closer than the sum of their radii) to eliminate even more spurious detection. We found that the gain was extremely limited ($0.01\%$ of the detected features suppressed) and that imprecision in positioning or in sizing caused valid particles to be discarded.

We note that the orientation of non-spherical particles or the deformation of soft objects can be monitored in principle by using the local Hessian matrix.

\subsection{Finite dilution}
The Gaussian (and $DoG$) response of a particle decays rapidly away from its surface (see Eq.~(\ref{eq:split_G}). We found that even at contact the position shift induced by a nearby particle was less than a tenth of a pixel (see Fig.~\ref{fig:perfect}c). However when two particles are closer than a few times the blurring width, their influence on each other cannot be neglected: the minima of the $DoG$ is effectively shifted toward smaller scales, leading to smaller radii if one uses Eq.~(\ref{eq:scale_dil}) out of the dilute context (see Fig.~\ref{fig:perfect}d).

The response of $N$ particles of radii $\lbrace R_i\rbrace$ can be superimposed at any scale $\sigma$ and at any point of space, in particular we define $DoG_i$ as the response at the centre of particle $i$.
\begin{equation}
DoG_i(\lbrace A_j\rbrace, \lbrace r_{ij}\rbrace, \lbrace R_j\rbrace, \sigma) = \sum_j A_j DoG(r=r_{ij}, R=R_j, \sigma),
\label{eq:superpos}
\end{equation}
where the function $DoG(r,R,\sigma)$ is the response of a particle of radius $R$ at distance $r$ from its centre; $r_{ij}$ is the distances between particles $i$ and $j$ and $\lbrace A_i\rbrace$ are the respective brightness of the particles. The \textsc{sift} algorithm yields $\lbrace \sigma_i^*\rbrace$ so that for all $i$, $DoG_i$ is minimum with respect to $\sigma$ at $\sigma_i^*$, thus differentiating Eq.~(\ref{eq:superpos}) with respect to $\sigma$:
\begin{equation}
\forall i,\quad \frac{\partial DoG_i}{\partial\sigma}(\lbrace A_j\rbrace, \lbrace r_{ij}\rbrace, \lbrace R_j\rbrace, \sigma_i^*) = 0.
\label{eq:DoG_min}
\end{equation}

The system defined by Eq.~(\ref{eq:DoG_min}) is non-linear with respect to $\lbrace R_j\rbrace$ but can be solved iteratively by Newton's method:
\begin{equation}
\left[ \frac{\partial^2 DoG_i}{\partial R_j\partial\sigma}\right] \times \left( \lbrace R_j\rbrace^{(k+1)} - \lbrace R_j\rbrace^{(k)} \right) = -\frac{\partial DoG_i}{\partial\sigma},
\label{eq:Newton}
\end{equation}
where the matrix and the right hand side are computed given $(\lbrace A_j\rbrace, \lbrace r_{ij}\rbrace, \sigma_i^*)$ and iteratively $\lbrace R_j\rbrace^{(k)}$, with the upper parenthesised index indicating the iteration rank. The results of Eq.~(\ref{eq:scale_dil}) are good starting values for the radii.

Using Eq.~(\ref{eq:superpos}), the elements of the matrix simplify to
\begin{equation}
\frac{\partial^2 DoG_i}{\partial R_j\partial\sigma} =  A_j \frac{\partial^2 DoG}{\partial R\partial\sigma}(r=r_{ij}, R=R_j^{(k)}, \sigma=\sigma_i^*).
\end{equation}

In principle, Eq.~(\ref{eq:Newton}) is a $N\times N$ system of equations. However the $DoG$ functions and its derivatives are rapidly decaying functions, thus the matrices are actually very sparse (about as many non-zero coefficients as particles in the first coordination shell), alleviating dramatically the computational burden when using sparse system solvers. Fig.~\ref{fig:perfect}d shows the result of such correction for two identical particles, where Eq.~(\ref{eq:Newton}) converges in a single iteration. In a many body case (see Fig.~\ref{fig:perfect}e) the convergence is reached in two iterations; however, the extremely low error of the dilute case is not totally recovered ($\approx 3\%$ relative error rather than $0.3\%$).

\subsection{Brightnesses}

To solve Eq.~(\ref{eq:Newton}), one needs knowledge of the brightnesses $\lbrace A_i\rbrace$. In a first approximation, they can be assumed equal to a constant, which allows to simplify them out. This is often a sensible approximation. Nevertheless, the particles in an experimental image are not uniformly bright due to synthesis imperfection (quantity of dye fixed by each particle) and photo bleaching. If one does not take into account the relative brightness of the particles, less bright particles will appear smaller.

A better approximation is to measure during the \textsc{sift} process the value of the $DoG$ response at the position and scale of each particle, \emph{i.e.} $DoG_i(\lbrace A_j\rbrace, \lbrace r_{ij}\rbrace, \lbrace R_j\rbrace, \sigma_i^*)$. Given the (iterative) values of the $\lbrace R_j\rbrace^{(k)}$, one can solve Eq.~(\ref{eq:superpos}) to get an iterative value of $\lbrace A_i\rbrace^{(k)}$. With respect to the brightnesses, Eq.~(\ref{eq:superpos}) is a linear system of $N$ equations with $N$ unknowns, thus directly solvable. It is also as sparse as Eq.~(\ref{eq:Newton}).

To sum up, the coefficients $\lbrace A_i\rbrace$ can be computed along with the radii in a joint iterative process:
\begin{algorithmic}
\State $\lbrace R_i\rbrace^{(0)} \xleftarrow{\text{Eq.~(\ref{eq:scale_dil})}} \lbrace \sigma_i^*\rbrace$
\Repeat 
	\State $\lbrace A_i\rbrace^{(k+1)} \xleftarrow{\text{Eq.~(\ref{eq:superpos})}} \lbrace DoG_i \rbrace, \lbrace R_i\rbrace^{(k)}, \lbrace \sigma_i^*\rbrace, \lbrace r_{ij}\rbrace$
	\State $\lbrace R_i\rbrace^{(k+1)} \xleftarrow{\text{Eq.~(\ref{eq:Newton})}} \lbrace R_i\rbrace^{(k)}, \lbrace A_i\rbrace^{(k+1)}, \lbrace \sigma_i^*\rbrace, \lbrace r_{ij}\rbrace$ 
\Until{convergence}
\end{algorithmic}

In our tests, we found that for both cases with and without the brightness determination this algorithm converges quickly in one or two iterations.

\section{Application to 3D confocal microscopy images}
\label{sec:confocal}

\subsection{Effect of a point spread function}
\begin{figure*}
\centering
\includegraphics{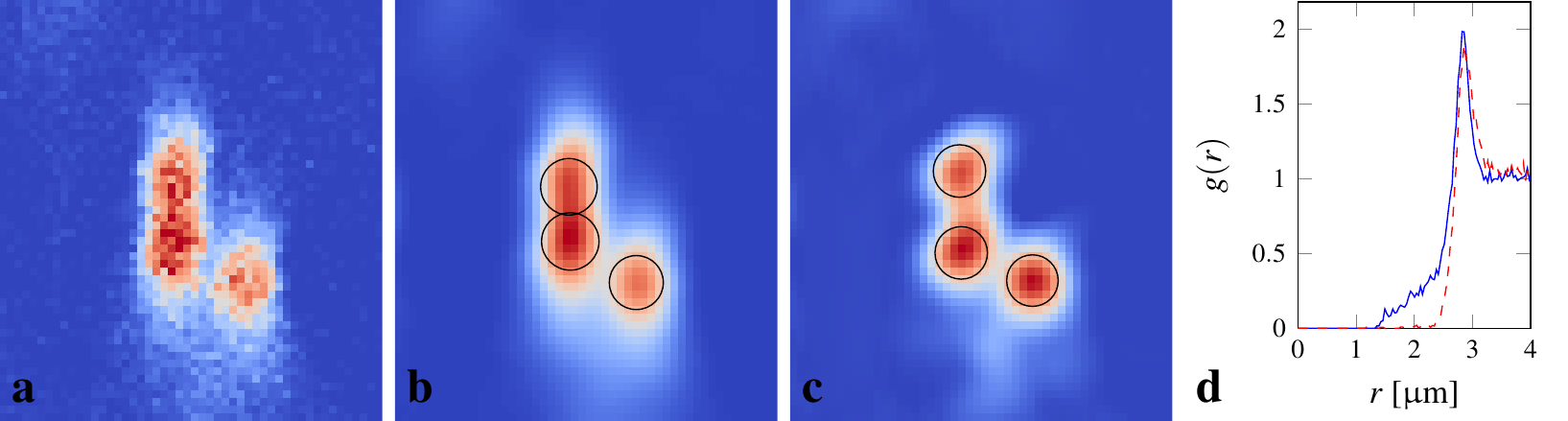}
	\caption{Deconvolution. Detail of the same $YZ$ slice of (a) original confocal image, (b) previous blurred by $\sigma_0=1.6$, (c) previous deconvolved by measured kernel. Circles indicate the tracked particles position and size when using whether (b) or (c) as first Gaussian layer $G_0$. All three centres are in the slice $\pm \unit{0.5}{px}$. (d) Radial distribution function of almost monodisperse sticky spheres localised (blue) without and (red dash) with deconvolution.}
	\label{fig:deconv}
\end{figure*}

Real images suffer from optical limitations, \emph{e.g.}, the point-spread function (\textsc{psf}) of the microscope. In particular, spherical particles observed by confocal microscopy appear elongated along $Z$. The influence of such anisotropic distortion on the Gaussian and $DoG$ responses is not trivial and most of the equations given in Appendix do not have analytical equivalents. In particular, the minimum of the $DoG$ response is found at larger values of $R/\sigma$, thus the naive use of the methods detailed in the previous section lead to overestimate particle sizes.

In addition, the overlap of neighbouring particle images is no longer negligible when the particles are aligned along $Z$. This leads to large imprecision in particle positions, especially in the case of anisotropic environment (\emph{e.g.}, isolated pair of particles, interface, colloidal gel). In Fig.~\ref{fig:deconv}d the radial distribution function of almost monodisperse colloids with short range attraction displays a spurious shoulder before its first peak, indicating that particles centres are often found closer than the sum of their hard core radii, as illustrated in Fig.~\ref{fig:deconv}b.

We found that these issues can be better addressed by pre-processing the images (as detailed below) rather than post processing the \textsc{sift} output \emph{via} analytical methods.

\subsection{Deconvolution}
The image $y$ acquired by a microscope can be expressed as 
\begin{equation}
y = x \star h + \epsilon,
\label{eq:psf}
\end{equation}
where $x$ is the perfect image, $h$ is the \textsc{psf} of the microscope and $\epsilon$ is the noise independent of both $x$ and $h$. The process of estimating $x$ from $y$ and some theoretical or measured expression of $h$ is called deconvolution. Deconvolution in the presence of noise is a difficult problem~\cite{Riad1986}. Hopefully here we do not need to reconstruct the original image, but only our first Gaussian blurred version of it, \emph{i.e.}, $G_0 = x \star g_{\sigma_0}$ starting from $y_0 = y \star g_{\sigma_0}$. Indeed, after a reasonable amount of blur in three dimensions, the noise can be neglected and we thus obtain:
\begin{equation}
y_0 \approx G_0 \star h,
\end{equation}
or in Fourier space
\begin{equation}
\mathcal{F}[y_0] = \mathcal{F}[G_0] \times \mathcal{F}[h]. 
\label{eq:Fourier_conv}
\end{equation}

Once $\mathcal{F}[h]$ is known the deconvolution reduces to a simple division in Fourier space. Let us measure $\mathcal{F}[h]$ in an isotropic system where we can write
\begin{eqnarray}
\left\langle \left|\mathcal{F}_X[x]\right|^2 \right\rangle  &= \left\langle \left|\mathcal{F}_Z[x]\right|^2 \right\rangle. 
\end{eqnarray}
Using Eq.~(\ref{eq:psf}) we obtain 
\begin{eqnarray}
\frac{\left\langle\left|\mathcal{F}_X[y]\right|^2 \right\rangle}{\left|\mathcal{F}_X[h]\right|^2}  = \frac{\left\langle \left|\mathcal{F}_Z[y]\right|^2\right\rangle}{\left|\mathcal{F}_Z[h]\right|^2}. 
\end{eqnarray}
Here $\mathcal{F}_X$ indicates the Fourier transform along axis $X$. In point scanning confocal imaging, the \textsc{psf} has negligible lobes along $X$ and $Y$ ($X$ only for line scanning), thus we have 
\begin{eqnarray}
\left|\mathcal{F}_Z[h]\right|^2 = \frac{\left\langle \left|\mathcal{F}_Z[y]\right|^2\right\rangle}{\left\langle\left|\mathcal{F}_X[y]\right|^2 \right\rangle}.
\end{eqnarray}
For example an Hermitian kernel (real valued spectrum) is 
\begin{eqnarray}
\mathcal{F}_Z[h] = \sqrt{\frac{\left\langle \left|\mathcal{F}_Z[y]\right|^2\right\rangle}{\left\langle\left|\mathcal{F}_X[y]\right|^2 \right\rangle}}.
\end{eqnarray}

Fig.~\ref{fig:deconv}b-c shows the particles localized from the original image and from the deconvolved image. Deconvolution mends both size overestimation and imprecision in $z$ coordinate. This also translates in the radial distribution function (Fig.~\ref{fig:deconv}d), where the spurious shoulder before the first peak disappears.

\subsection{From optical to hard-core radius}

The method exposed above relies on the uniformity of the brightness within each particle to extract optical radii that is a good approximation to the physical (hard core) radii. However a colloidal particle often shows a smooth intensity profile in itself under the microscope, less bright at the edge than at the centre of the particle. This can be due to inhomogeneous fixation of the dye during the colloid synthesis, but the bottom line is the in-plane \textsc{psf} not corrected for in our above deconvolution method.

Under such conditions, the particles are detected smaller than their expected sizes. The real to optical size ratio has to be set using our knowledge of the sample. For example one may use the position of the first peak of the $g(r)$ to measure the average hard-core diameter. In general the real to optical size ratio may depend on the size of the particles and may evolve in time (\emph{i.e.}, because of photobleaching). We present below a detailed analysis of these issues for a case system. 

We also note that darker edges lead to smaller influence of a particle on the $DoG$ response of its neighbours. A smaller optical radius makes the particles optically further from each other. Accordingly, we found that finite dilution corrections were less important in real images than in our synthetic test images. For the same reason, the scaling factor must be measured and applied after --- \emph{not} before --- finite dilution corrections.

\section{Revealing polydispersity effects on structural orderings in a glassy colloidal hard sphere liquid}
\label{sec:yon6}

Here we apply our tracking method to access couplings between the spatial particle size distribution and local structural ordering 
in a supercooled colloidal liquid with size polydispersity.  

\subsection{Experimental}
We used \textsc{pmma} (poly(methyl methacrylate)) colloids sterically stabilized with methacryloxypropyl terminated \textsc{pdms} (poly(dimethyl siloxane)) and fluorescently labelled with rhodamine isothiocyanate chemically bonded to the \textsc{pmma}. The colloids were suspended in a solvent mixture of cis-decalin and cyclohexyl-bromide for both optical index and density matching. To screen any (weak) electrostatic interactions, we dissolved tetrabutylammonium bromide salt, to a concentration of \unit{300}{\nano\mole\per\liter}~\citep{royall2005}. The estimated Debye screening length is \unit{13}{\nano\metre}, well below the length scale of the colloids that can be considered as hard spheres. The \unit{8}{bit} graylevel data was collected on a Leica SP5 confocal microscope, using \unit{532}{\nano\meter} laser excitation and voxel size of $(\unit{283}{\nano\metre})^2\times\unit{293}{\nano\metre}$.

To realize a more precise density matching, the temperature was controlled on both stage and objective lens. This setup allows us to alter the buoyancy simply by a temperature change, thus to set the effective gravity upward, downward or almost null. After careful shear melting, the sample was filled into a $\unit{100}{\micro\metre}\times\unit{1}{\milli\metre}$ capillary (Vitrocom) and set on the microscope stage. We then spent a few days to find the temperature corresponding to exact density matching (within \unit{0.1}{K}). This waiting time was enough for crystallites to form on both top and and bottom walls. Then, we heat up our sample by a few degrees compared to the density-matching temperature in order to make the colloids heavier than the solvent. At the top of the sample (close to the objective lens and thus allowing much clear imaging) the volume fraction drops and all crystallites melt. Finally, we set back the thermostat to the density-matching temperature, allowing the top of the sample to slowly return to its supercooled state. We could then observe the heterogeneous nucleation from the beginning. We tracked the sample from top to bottom and we are thus sure that the crystallite actually form at the wall and do not come from the rest of the sample. Namely, crystallisation in this sample is caused only by heterogeneous nucleation on the substrates, presumably due to the wall-induced 
enhancement of crystal-like bond orientational ordering \cite{watanabe2011}.  

\subsection{Global size distribution}

We localise the particles using our method with a preblur of only $\sigma_0=1.0$ (see Section \ref{sec:blur}) to be able to detect the smaller particles ($R_{min}\approx \unit{2}{px}$) without expensive oversampling. We checked that choosing higher $\sigma_0$ truncates the size distribution but has no other effect on its shape. After removing half-overlapping features, we applied a single iteration of finite dilution corrections for both intensities and radii to obtain the optical radius of each particle.

\begin{figure}[h]
\centering
\includegraphics{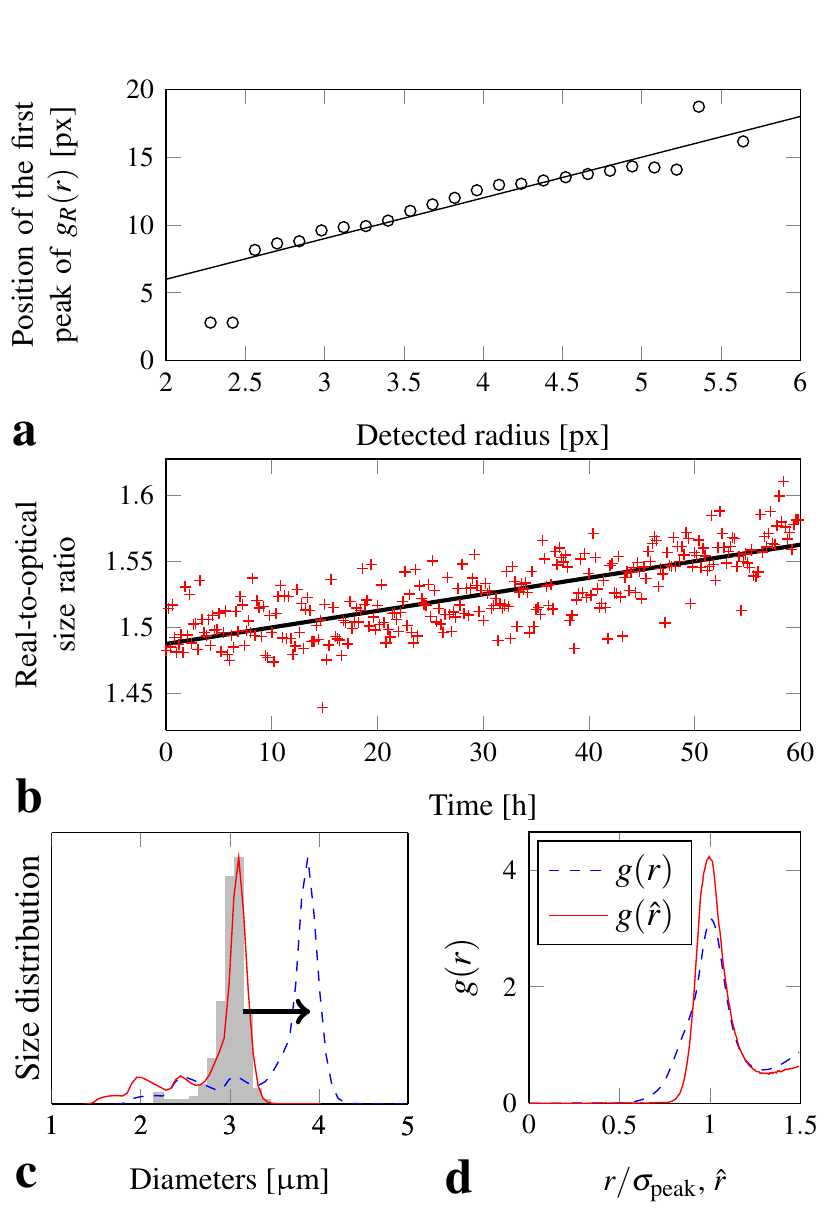}
	\caption{\emph{In situ} sizing of colloids in a glass. (a) Position of the first peak of the partial $g_R(r)$ function of the optical radius $R$. Solid line corresponds to a real-to-optical size ratio of $1.5$. (b) Time dependence of the real-to-optical size ratio. Solid line is a linear fit. (c) Size distribution estimated by our algorithm (dashed line). Comparison with the estimation from \textsc{sem} of only $140$ dry particles (steps) is possible once $25\%$ of swelling is taken into account (full line). (d) First peak of the radial distribution function with (full line) and without (dashed) the individual sizes data.}
	\label{fig:sizing}
\end{figure}

We then estimate the hard-core diameter function of the optical radius $R$ by locating the first peak of the partial radial distribution function $g_R(r)$ of the particles having $R_i$ close to $R$. As shown in Fig.~\ref{fig:sizing}a, the real to optical size ratio is around $1.5$ and is rather constant respective to $R$, thus a single overall scaling factor is enough. It can be determined by a single partial $g_R(r)$ with $R$ near the peak of the size distribution, or by locating the peak of $g(\hat{r})$, with $\hat{r}_{ij} = r_{ij}/(R_i+R_j)$. We found that the real-to-optical size ratio was increasing with time due to photobleaching (see Fig.~\ref{fig:sizing}b). We fit this increase by a linear relation and applied the resulting time-dependent ratio to obtain the real size of each particles.

The consistency of our method can be checked by constructing the radial distribution function $g(r)$. In monodisperse hard spheres, the $g(r)$ should have a sharp first peak at $r=2R$ corresponding to hard core contact. Polydispersity implies hard core contacts at various $r$ and thus broadens the peak. One can recover a sharp peak by constructing $g(\hat{r})$ (this time $\hat{r}$ is scaled by the real radii, not the optical ones). In Fig.~\ref{fig:sizing}d we successfully used the sizes measured by our method to rectify the first peak.

Fig.~\ref{fig:sizing}c shows the size distribution only $140$ dry colloids measured on high resolution scanning electron microscopy (\textsc{sem}) images. We also show the size distribution obtained by our \emph{in situ} measurements ($\sim 1.7\times 10^6$ instantaneous sizing). The main peak of the later compares well with the former once a solvent swelling of $25\%$ in radius is taken into account. However the small sampling of the SEM measurements completely misses the tail toward small sizes featuring two low peaks that probably correspond to secondary and ternary nucleation during the synthesis of the colloids~\cite{bosma2002,Poon2012}.

From our data we compute an overall volume fraction of $0.60$, almost constant during the experiment. The polydispersity estimated from \textsc{sem} data is $6.2\%$. This is coherent with a fit of the main peak of our data ($6.9\%$), however the polydispersity of the whole distribution including the tail toward small sizes is $14.8\%$. We will see below how such a complex size distribution affects the physical behaviour of the system.

\subsection{Polydispersity and structural heterogeneity in a supercooled colloidal liquid}

\begin{figure}[h]
	\centering
	\includegraphics{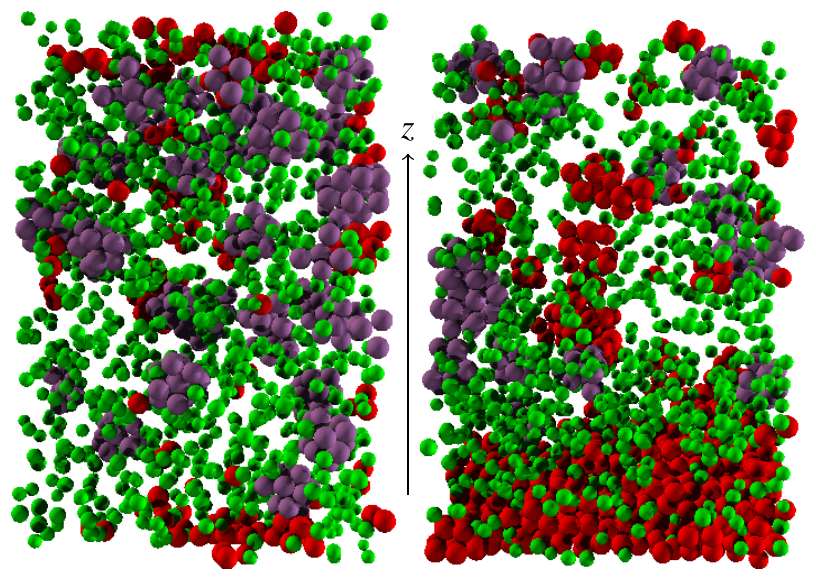}
	\caption{Structure visualisation at $t=0$ (left) and $t=\unit{40}{\hour}$ (right). Small particles ($R<\unit{1.5}{\micro\metre}$) are shown in green, crystal-like ordered particles ($Q_6>0.25$) in red and icosahedral particles and their neighbours in purple. Other particles are not shown.}
	\label{fig:mrco_ico_small}
\end{figure}

Hard sphere supercooled liquids and glasses contain both medium ranged crystal-like bond orientational ordering (\textsc{mrco}) and local icosahedral ordering~\cite{Leocmach2012}. The later acts as a frustration against the expansion of the former and thus against crystallisation. Both kind of structures can be detected using bond orientational order (\textsc{boo}) introduced by \citet{steinhardt1983boo} (see Fig.~\ref{fig:mrco_ico_small}). As we have described elsewhere~\cite{Leocmach2012}, crystal-like bond ordering is well described by the scalar bond order parameter $Q_6$, and the icosahedral bond ordering by $w_6$. \textsc{fcc} or \textsc{hcp} crystals under thermal vibrations typically have $Q_6>0.4$~\cite{Lechner2008}. We distinguish \textsc{mrco} from liquid structures by a threshold at $Q_6=0.25$. The perfect 13 particles icosahedron is the minimum of $w_6$, at negative values. We consider that a neighbourhood is icosahedral when $w_6<-0.023$.

\begin{figure}
	\centering
	\includegraphics{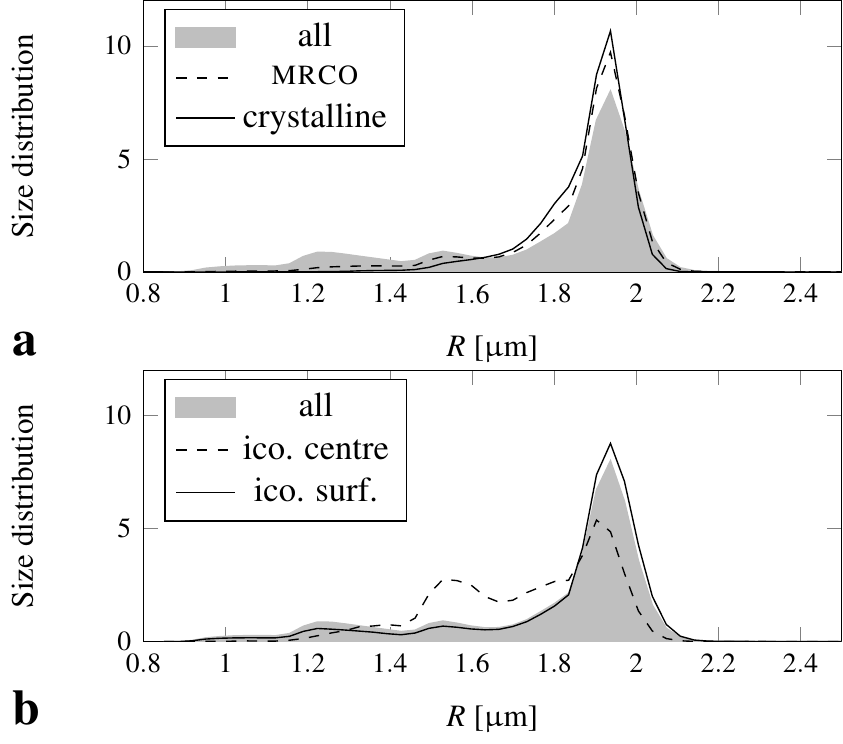}
	\caption{Size distribution within local structures in a supercooled hard sphere colloidal liquid. (a) With increasing local crystalline bond orientational order: all particles, MRCO particles ($0.4>Q_6>0.25$) and crystalline particles ($Q_6>0.4$). Crystalline ordered regions are less polydisperse. (b) Particles at the centre or on the surface of icosahedra ($w_6<-0.023$). A small central particle with large surface particles tend to stabilize icosahedra.}
	\label{fig:size_struc}
\end{figure}

In Fig.~\ref{fig:size_struc}, we compare the overall size distribution to the size distribution within each remarkable structure. The roles of the particle at the centre of an icosahedron and of the particles surrounding it are asymmetric. The size distribution of the particles at the surface of icosahedra is almost identical to the overall size distribution, however the size distribution of the particles at the center of icosahedra features a second peak at radii about $80\%$ smaller than the main peak. A centre to surface size ratio near $0.8$ seems to stabilize icosahedral order. This is consistent with a recent simulation study by Shimono and Onodera \cite{shimono2012icosahedral}.  

We note that the tail of small particles characteristic of our overall size distribution is less pronounced in \textsc{mrco} and almost disappears in well-formed crystals (Fig.~\ref{fig:size_struc}a). The width of the main peak of the distribution does not change significantly upon crystal-like ordering. This means that a small amount of polydispersity is acceptable inside \textsc{mrco}, coherently with the results for the (defective) crystalline lattice by \citet{Fasolo2003}. However \textsc{mrco} cannot appear where markedly smaller particles are present. We also checked the difference between \textsc{fcc} ($w_4<0$) and \textsc{hcp} ($w_4>0$) to find that they have both exactly the same size distribution (not shown). This suggests that the size distribution controls the formation of \textsc{mrco} but has no influence on its symmetry (probably, the crystal 
polymorph). 
We stress that this coupling of \textsc{mrco} and icosahedral ordering with the spatial distribution of particle sizes does not imply fractionation, but rather suggests
that a rather homogeneous size distribution tends to help or stabilize \textsc{mrco} and a slightly smaller size of a particle also stabilizes the formation of icosahedral order 
around it. We will see below that the fractionation appends later in the crystallisation process.

\subsection{Polydispersity and heterogeneous crystallisation in a supercooled colloidal liquid}

\begin{figure}
	\centering
	\includegraphics{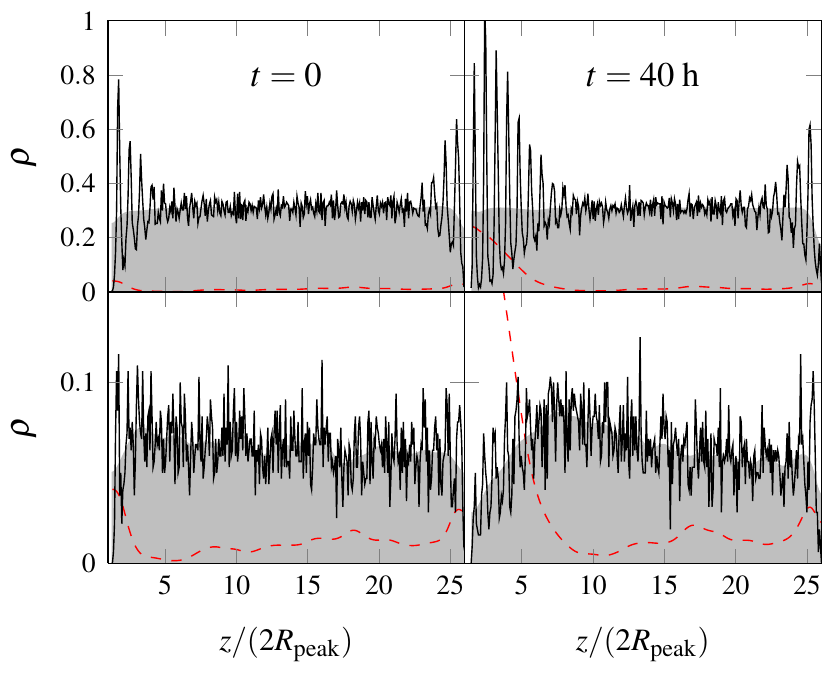}
	\caption{Instantaneous density profiles at $t=0$ (left) and $t=\unit{40}{\hour}$ (right). Top: large particles ($R>\unit{1.5}{\micro\metre}$) without smoothing (black line) and with a Gaussian smoothing of width $2R_\text{peak}$ (gray area). Bottom: small particles ($R<\unit{1.5}{\micro\metre}$) with the same color code. 
For all figures, the red dashed curve is the smoothed density of ordered particles ($Q_6>0.25$) irrespective of their size.}
	\label{fig:profiles}
\end{figure}

As explained above, we triggered heterogeneous crystal nucleation on the top wall of our container. At each time step we can construct a $z$-dependent density profile for any species of interest. In Fig.~\ref{fig:profiles} we show at two time steps the density profiles of large particles ($R>\unit{1.5}{\micro\metre}$), small particles ($R<\unit{1.5}{\micro\metre}$) and particles with medium to high crystalline order ($Q_6>0.25$). Icosahedral particles (not shown) are an order of magnitude fewer than small particles (see Fig.~\ref{fig:mrco_ico_small}) and are basically irrelevant. Near the walls, the density profiles show oscillations characteristic of layering 
(see, \emph{e.g.}, Refs. \cite{watanabe2011,kob3}), whose wavelength corresponds to the peak of the size distribution (the `majority species'). We note that the small particles also shows layering on the same wavelength, although with a smaller amplitude.

\begin{figure}
	\centering
	\includegraphics{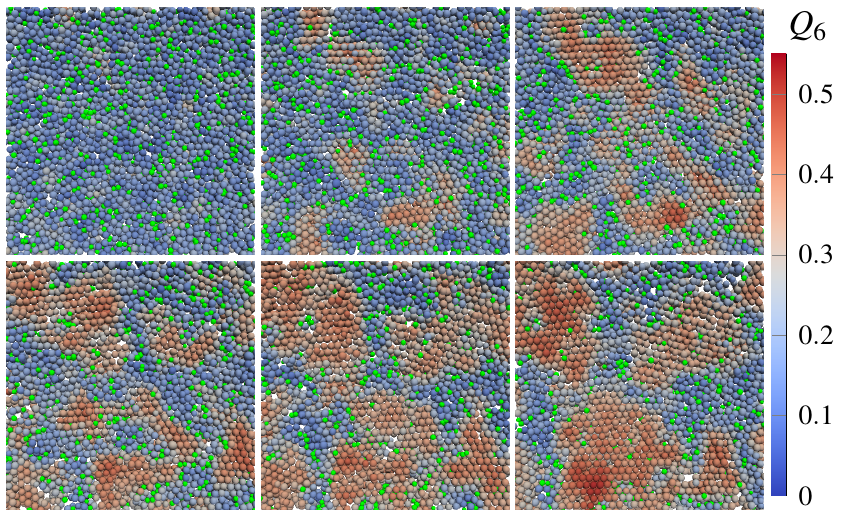}
	\caption{Heterogeneous nucleation. Successive computer reconstruction of the first three layers from the wall, every \unit{12}{\hour}, seen from the fluid side. Particles sizes are according to our method. Small ($R<\unit{1.5}{\micro\metre}$) particles are shown in green. Other particles are coloured according to their crystal-like order. Note how the small particles are expelled from the forming crystallites to concentrate in the grain boundaries.}
	\label{fig:small_expelled}
\end{figure}

Once the short oscillations removed via a Gaussian smoothing of width $2R_\text{peak}$, one can see that the density of large particles is constant from top to bottom of the sample (the slight dip on the edges is due to a small misalignment between our $z$ axis and the normal of the wall). It confirms the accuracy of our density matching but also stresses that crystallisation takes place without an increase in density. The smoothed density profile of small particles starts almost flat. However the small particles are expelled from the forming crystal, and as the crystalline front propagates it pushes in front of itself a growing ridge of small particles, very apparent after $\unit{40}{\hour}$ in both Fig.~\ref{fig:profiles} and Fig.~\ref{fig:mrco_ico_small}.

We confirmed this view by looking at the behaviour of small particles and crystalline order in a constant plain (Fig.~\ref{fig:small_expelled} and Supplementary Movie 1). Before crystallisation, small particles are scattered randomly and \textsc{mrco} develop where few very small particles are present. Then, crystallisation proceeds from the \textsc{mrco}, pushing away the small particles that accumulate in the grain boundaries. As a consequence, heterogeneous crystallisation do not proceed via a spatially homogeneous layer-by-layer growth, as observed at low polydispersity~\cite{Sandomirski2011}, but by many nucleation-growth events reflecting the initial fluctuations of the \textsc{mrco}. This dynamics is reminiscent of the homogeneous nucleation case~\cite{Kawasaki2010c,Russo2012,Russo2012b} but enhanced by the wetting of the wall by crystalline bond orientational order \cite{watanabe2011}.

To sum up, the small particles affect the growth of the crystal by coupling ordering to diffusion. Without the small particles, moving grain boundaries implies moving particles only locally and thus propagation is easy. With very small particles, moving grain boundaries implies moving them by the same amount by diffusion. In the same way, the crystallisation front has to push the small particles in front of it, effectively increasing the polydispersity in the melt and thus slowing its propagation. Indeed, weeks after preparation our samples was still only partially crystallised and highly polycrystalline. We note that this coupling is due to the conserved nature of the species `small particles'. Non-conserved species defined by local structural ordering such as icosahedra cannot produce such a coupling.

\section{Conclusion}
\label{sec:conclusion}

We have presented a new method to extract the coordinates of particles of very diverse sizes even when they are very close (at contact) to each other. This opens new experimental possibilities in soft matter systems where the particles are polydisperse (\emph{e.g.}, colloids, emulsions, and granular particles) or even changing size (\emph{e.g.}, phase separation dynamics). Our method also reliably extracts the size of individual particle, which allows a \emph{in situ} analysis of the consequences of naturally occurring size distributions (beyond monodisperse, bidisperse or Gaussian distribution).

We showed that in the case of colloidal hard sphere, a size distribution with a main peak and a tail toward small sizes induces different behaviours than what would be expected from a Gaussian size distribution. Our method allowed us to measure this distribution properly and to look into the interactions between size distribution, local ordering and heterogeneous crystal nucleation in a glass. We found that the presence of a long tail toward small sizes in the size distribution, was leading to highly polycrystalline materials that may be of interest for engineering purposes.
We hope that this novel particle tracking method with a capability of particle-size determination will be applied to a wide field in soft matter physics. 

\section*{Acknowledgements}
We are grateful to Koshi Hasatani, Anthony Genot and Yannick Rondelez for providing the image of Fig.~\ref{fig:applications}a; to Ken-ichiro Murata for providing the image of Fig.~\ref{fig:applications}b; to John Russo for providing us with simulation data of dense hard spheres, which are used in Fig.~\ref{fig:perfect}e; to Hideyo Tsurusawa for the synthesis of the colloid and the confocal data of Fig.~\ref{fig:deconv}; and to Rebecca Rice and C. Paddy Royall for \textsc{sem} measurements of the size distribution of dried particles. 
This work was partially supported by a grant-in-aid from the 
Ministry of Education, Culture, Sports, Science and Technology, Japan and also by 
Aihara Project, the FIRST program from JSPS, initiated by CSTP.

\appendix
\section*{Appendix: Gaussian response of a binary ball}
\label{sec:gaussian_vs_ball}

A binary ball $\mathcal{B}_R$ of radius $R$ is convolved by a Gaussian kernel of width $\sigma$. The response at a distance $r$ (along z-axis without loss of generality) is

\begin{align}
G(r,R,\sigma) =& \frac{1}{(2\pi)^{3/2}\sigma^3} \int_{\mathcal{B}_R}{e^{ -\frac{(z-r)^2+y^2+x^2}{2\sigma^2}} dx dy dz }
\intertext{or in spherical coordinates $(\rho, \theta, \phi)$ after integration along $\phi$}
=& \frac{e^{-\frac{r^2}{2\sigma^2}}}{\sqrt{2\pi}\sigma^3} \int_0^R \rho^2 e^{-\frac{\rho^2}{2\sigma^2}} d\rho \int_0^\pi \sin\theta e^{\frac{\rho r}{\sigma^2}\cos\theta} d\theta \\
 =& \frac{1}{r\sigma\sqrt{2\pi}}\left[ \int_0^R \rho e^{-\frac{(\rho-r)^2}{2\sigma^2}}d\rho - \int_0^R \rho e^{-\frac{(\rho+r)^2}{2\sigma^2}}d\rho \right] 
\intertext{this can be integrated using the error function $\erf$}
G(r,R,\sigma) &= \frac{1}{2}\left( g(r,R,\sigma) + g(-r,R,\sigma)\right)
\label{eq:split_G}
\intertext{with}
g(r,R,\sigma) &= \erf\left(\frac{R+r}{\sigma\sqrt{2}}\right) + \sqrt{\frac{2}{\pi}}\frac{\sigma}{r}e^{-\frac{(R+r)^2}{2\sigma^2}}
\intertext{At the centre of the ball, Eq.~(\ref{eq:split_G}) reduces to}
G(r=0,R,\sigma) &= \erf\left(\frac{R}{\sigma\sqrt{2}}\right) - \sqrt{\frac{2}{\pi}}\frac{R}{\sigma}e^{-\frac{R^2}{2\sigma^2}}
\end{align}

In addition, we compute the following useful partial derivatives
\begin{align}
\frac{\partial g}{\partial \sigma}(r,R,\sigma) &= \frac{\left( R^2+r R+\sigma^2\right)}{r\sigma^2\sqrt{2\pi}} e^{-\frac{(R+r)^2}{2\sigma^2}} \\
\frac{\partial^2 g}{\partial \sigma \partial R}(r,R,\sigma) &= -R \frac{(R+r)^2-\sigma^2}{r\sigma^4\sqrt{2\pi}} e^{-\frac{(R+r)^2}{2\sigma^2}}\\
\frac{\partial G}{\partial \sigma}(r=0,R,\sigma) &= -\sqrt{\frac{2}{\pi}} \frac{R^3}{\sigma^4} e^{-\frac{R^2}{2\sigma^2}} \label{eq:Gc_ds}\\
\frac{\partial^2 G}{\partial \sigma \partial R}(r=0,R,\sigma) &= \sqrt{\frac{2}{\pi}} R^2 \frac{R^2-3\sigma^2}{\sigma^6} e^{-\frac{R^2}{2\sigma^2}}
\end{align}

The difference of Gaussians response is 
\begin{align}
DoG(r,R,\sigma,\alpha) &= G(r,R,\alpha\sigma)-G(r,R,\sigma)
\intertext{and has the partial derivative relative to the scale}
\frac{\partial DoG}{\partial \sigma}(r,R,\sigma,\alpha) &= \alpha \frac{\partial G}{\partial \sigma}(r,R,\alpha\sigma) - \frac{\partial G}{\partial \sigma}(r,R,\sigma) \label{eq:DoG_ds}
\end{align}
When the difference of Gaussians is minimum at the centre of the ball, we combine Eq.~(\ref{eq:Gc_ds}) and Eq.~(\ref{eq:DoG_ds}) to get
\begin{equation}
\frac{R}{\sigma\sqrt{2}} = \sqrt{\frac{3\ln \alpha}{1-\alpha^{-2}}} \label{eq:R_dilute}
\end{equation}

One can show that the factor $3$ in the radical is the dimensionality of the space, thus Eq.~(\ref{eq:R_dilute}) can be generalised to other dimensions.

\end{document}